\documentstyle[12pt]{article}
\newcommand{\beq}{\begin{equation} }
\newcommand{\eeq}{\end{equation} }
\newcommand{\ue}{ U_\epsilon }
\newcommand{\uex}{U_\epsilon^X }
\newcommand{\spec}{\sigma(\uex)}
\def \C{{\bf C}}
\def \N{{\bf N}}
\def \Z{{\bf Z}}
\def \R{{\bf R}}
\def \E{{\cal E}}
\def \H{{\cal H}}
\def \F{{\cal F}}
\def \D{\overline{D}}
\def \epsilon{\varepsilon}

\begin{document}

\begin{center}
{\Large {\bf Resonances of the cusp family}}
\vskip0.5cm \noindent
{\bf I. Antoniou$^{1,2}$, S. A.
Shkarin$^{1,3}$, E.  Yarevsky$^{1,4}$}
\end{center}

\vskip0.3cm \noindent
$^1${\it International Solvay Institutes for Physics
and Chemistry, Campus Plaine ULB C.P.231, Bd. du Triomphe,
Brussels 1050, Belgium} \newline
$^2${\it Department of Mathematics, Aristotle University of 
Thessaloniki, 54006, Greece} \newline
$^3${\it Moscow State University, Dept. of Mathematics and Mechanics,
Vorobjovy Gory, Moscow, 119899, Russia} \newline
$^4${\it Laboratory of Complex Systems Theory, Institute for Physics,
St.Petersburg State University, Uljanovskaya 1, Petrodvoretz,
St. Petersburg 198904, Russia}

\begin {abstract}
We study a family of chaotic maps with limit cases the tent map and
the cusp map (the cusp family). We discuss the spectral properties
of the corresponding Frobenius--Perron operator in different function
spaces including spaces of analytic functions.
A numerical study of the eigenvalues and eigenfunctions is performed.
\end {abstract}


\section {Introduction}

Resonances of dynamical systems play important role
in the study of the decay of correlations and are manifestations
of the statistical properties of chaotic systems. Hence
it is not surprising that they are studied very
intensively. We refer readers to recent reviews on this vast
subject~\cite {ergodic,baladi}. Resonances appear also in the 
generalized spectra \cite {AntTas,ADMK,AntShk} of the evolution 
operators \cite {logistic,sinai} of chaotic maps.

The theory of resonances has been
recently developed in terms of locally convex topological vector
spaces~\cite {AntTas,ADMK,AntShk}. This reflects the fact that
dynamical systems are defined in terms of the space of observables and
the evolution law. For different classes of observables the same 
evolution law may have different resonances i.e.
different rates of approach to equilibrium. However once the class of
observables is chosen the resonance structure is
unique~\cite {AntShk,ASS}.  Therefore we have proposed~\cite
{AntShk,ASS,tentres} that physical equivalence should reflect
identical physical properties i.e. rates of decay of
correlations.

For many classes of maps, e.g. expanding maps, there exist some
exact results about existence of resonances and their
estimations~\cite {baladi}.  However, for more complicated
maps each case needs a separate consideration and results
are sparse. Their study has attracted a lot of interest.

For example, the so-called cusp map~\cite{cuspdef}
$$
F: [-1,1] \to [-1,1], \quad {\rm where} \quad F(x)=1-2\sqrt{|x|}
$$
is an approximation of the
Poincar\'e section of the Lorenz
attractor~\cite {lorenz,attractor}.
The absolutely continuous invariant probability
measure of the cusp map has density
$$
\rho(x)={1-x \over 2}.
$$
The cusp map is a limit case of the cusp family
\cite{family,kaufmann,correl}:
\beq
\begin{array}{l}
F_\epsilon : [-1,1] \rightarrow [-1,1],\qquad
\epsilon\in[0,1/2],\quad {\rm where}
\\
\displaystyle
F_\epsilon(x) = {1-\sqrt{1-4\epsilon(1-\epsilon-2|x|)} \over 2\epsilon}
{\rm \ \ for\ \ }\epsilon\in(0,1/2],
\\
\displaystyle
F_0(x)=\lim_{\epsilon\downarrow0}F_\epsilon(x)=1-2|x|.
\end{array}
\label {familydefinition}
\eeq
The map with $\epsilon=0$ is the well-known
tent map~\cite{tentdef} while the map with $\epsilon=1/2$
is the cusp map~\cite{cuspdef}.

Each map in the cusp family is an exact system. For
$\epsilon \neq 1/2$ it follows directly from theorem~4 in
\S~8, chap.~10 of~\cite{sinai}, which gives sufficient
conditions of exactness for piecewise monotonic maps. 
The exactness of the cusp map has been hinted by
Hemmer~\cite{cuspdef} referring to the work of Lasota and
Yorke~\cite{lasota}. The proposed hint seems to be
irrelevant as the cusp map has a parabolic fixed point.

For the cusp map one should consider the so-called
induced map \cite {induced1,induced2,induced3} on the segment
$[\sqrt8-3,3-\sqrt8]$. This map satisfies the conditions of
the above mentioned theorem and
therefore is exact. Since the exactness for a map and
its induced map are equivalent \cite {induced1,induced2,induced3}, 
we obtain the exactness of the cusp map.

The unique absolutely continuous Borel invariant probability 
measure $\mu_\epsilon$ for the cusp family $F_\epsilon$ has
density~\cite {family}
\beq
\rho_\epsilon(x)=\frac{1}{2}-\epsilon x.
\label {density}
\eeq

The statistical analysis of dynamical systems is based on
the Koopman and the Frobenius--Perron operators. The Koopman
operator of a measurable map $S:Y\to Y$, where $(Y,\F)$ is
a measurable space, acts on functions $f:Y\to\C$:
$$
Vf(x)=f(Sx).
$$
The Frobenius--Perron operator (F.P.O.) $U$ is defined with respect
to a probability reference measure $\nu$ on $(Y,\F)$. For
$1\leq p\leq\infty$ the F.P.O.
$U:L_p(Y,\F,\nu)\to L_p(Y,\F,\nu)$ is the dual of the
operator $V:L_q(Y,\F,\nu)\to L_q(Y,\F,\nu)$, where
$\frac1p+\frac1q=1$:
$$
(U\rho|f)=(\rho|Vf),\qquad(\rho|f)=\int\nu(dy)\,
\overline{\rho(y)}f(y).
$$

In case of an exact endomorphism $S$ on a segment
$Y=[a,b]$ one usually use either the normalized Lebesgue
measure or the invariant absolutely continuous probability
measure as the reference measure. In both cases 1 is an
eigenvalue of the F.P.O. However, in the
first case the corresponding eigenfunction is the density of
the invariant measure; in the second case the corresponding
eigenfunction is constant {\bf 1}. In our paper we use the
invariant measure as the reference one.

The Frobenius--Perron operator $\ue $
of $F_\epsilon$ with respect to
the invariant
measure $\mu_\epsilon$ is
\beq
\ue \rho(x) = \left(\frac{1}{2}-\epsilon a_\epsilon(x)\right)
\rho(a_\epsilon(x))
+\left(\frac{1}{2}+\epsilon a_\epsilon(x)\right) \rho(-a_\epsilon(x)),
\label{froben}
\eeq
where
$$
a_\epsilon (x) = {1-x \over 2} -{\epsilon \over 2}(1-x^2).
$$

The objective of this paper is to study the resonances of the cusp
family~(\ref{familydefinition}).
In Section~2 we present some definitions and results for the spectral
theory of operators necessary for the study of the F.P.O. of the cusp
family.
In Section~3 we present results about the spectral properties of the
F.P.O. generated by this family in different function spaces.
In Section~4 we analyze the spectral properties in spaces of analytic
functions.
In order to analyze the eigenvalues and eigenfunctions of the cusp family,
we perform in Section~5 a numerical study. We show that the cusp family does
not have spectrum in the form $r^n$, where $n \in \N$, $r \in \R$, 
in the space of analytic functions, at least in the
vicinity of the tent map. We analyze the behavior of the eigenvalues in the
vicinity of the cusp map. The behavior of the eigenfunctions is also
discussed.

\section {Normal points of linear operators}

Let $A$ be a linear continuous operator in a locally convex
topological linear
space $E$. The point $z\in\C$ is said to be {\it regular} if
$A-zI$ has continuous  inverse
(here and below $I$ is the identity operator).
The set of all nonregular points is the spectrum of $A$,
denoted as $\sigma(A)$. The point $z\in\C$ is said to be a
{\it normal point} \cite{g-k} if $E$ admits a
decomposition into a topological direct sum \cite{rob} of
two closed linear subspaces
\begin{equation}
E=E_0\oplus E_1
\label{sum}
\end{equation}
such that $E_0$ is finite dimensional,
$A(E_j)\subseteq E_j$ for $j\in\{0,1\}$,
$(A-zI)\bigr|_{E_1}:E_1\to E_1$ has continuous inverse
and there exists $n\in\N$ such that
$(A-zI)^n(E_0)=\{0\}$.

Evidently the point $z$ is regular if and only if it is normal
and $E_0=\{0\}$. A normal point for which $E_0\neq\{0\}$ is
called a {\it normal eigenvalue}.

It is well-known \cite{g-k,ed} that for any normal
point $z$ the decomposition (\ref{sum}) is unique.
Moreover, the monotonic sequences of spaces ker$\,(A-zI)^n$
and $(A-zI)^n(E)$ stabilize and
\begin{equation}
E_0=\bigcup_{n=1}^\infty\hbox{ker}\,(A-zI)^n,
\quad E_1=\bigcap_{n=1}^\infty(A-zI)^n(E).
\label{xo}
\end{equation}

For a normal point $z$ we denote
\begin{equation}
\E(z,A)=E_0.
\label{egs}
\end{equation}
Note that if $z$ is regular then $\E(z,A)=\{0\}$.
According to (\ref{xo}) the finite dimensional space
$\E(z,A)$ is spanned by the eigenvectors and the principal vectors
of $A$ associated to the eigenvalue $z$. In the case $\E(z,A)\neq\{0\}$
the dimension of $\E(z,A)$ is the {\it multiplicity} of the
normal eigenvalue $z$.

If the spectrum of $A$ is either finite or is a sequence
converging to 0 and any non-zero element of $\sigma(A)$
is a normal eigenvalue of $A$ (this happens
e.g. for any compact operator on a Banach space \cite{ed}),
we can relabel the spectrum $\sigma(A)$ as a sequence
$$
z_n(A),\qquad n=0,1,2,\dots
$$
so that the following conditions are fulfilled:
\begin {equation}
\begin {array}{l}
1)\ \  |z_{n+1}(A)|\leq|z_n(A)|\hbox{\ \ for all\ \ }
n\in\Z_+,
\\
2)\ \ \hbox{if\ \ }z_n(A)\neq0\hbox{\ \ then\ \ }
z_n(A)\in\sigma(A),
\\
3)\ \ \hbox{if\ \ }z\in\sigma(A)\setminus\{0\}
\hbox{\ \ then\ \ }|\{n\in\Z_+:z_n(A)=z\}|=
\hbox{dim}\,\E(z,A)
\\
4)\ \ \hbox{if\ \ }|z_n(A)|=|z_{n+1}(A)|\hbox{\
\ then\ \ }\hbox{arg}\,z_n(A)<\hbox{arg}
\,z_{n+1}(A),
\end {array}
\label{lan}
\end {equation}
where arg$\,z\in(-\pi,\pi]$ is the argument of the complex
number $z$.

\section {Spectral properties of the Frobenius--Perron
operator in $L_p$ and $C^k$}

Let us introduce the following notation:
\begin{equation}
\D(a,q)=\{z\in\C:|z-a| \leq q\}, \quad
D(a,q)=\{z\in\C:|z-a| < q\},
\label{D}
\end{equation}

For any $p\in[1,+\infty]$ we denote the Hardy space in the
disk $D(a,q)$ by $\H^p(a,q)$, i.e. $\H^p(a,q)$ is the space
of holomorphic functions $f:D(a,q)\to\C$, which belong
to $L^p(D(a,q
))$ with respect to the Lebesgue measure. We endow
this space with the $L_p$-norm.

The operator $\uex: X\to X$ is the restriction of $\ue$ to
a locally convex function space $X$ such that
$\ue(X)\subseteq X$.  The spectrum of the operator $\uex$
is denoted by $\spec$.

{\bf Proposition 1.} \sl Let $\epsilon\in[0,1/2]$, $X$ be
either the Banach space $C[-1,1]$ or
$L_p([-1,1],\mu_\epsilon)$. Then the spectrum $\spec$
coincides with the closed unit disk $\D(1)$. Moreover, any
$z$ from the open unit disk $D(1)$ is an eigenvalue of
$\uex$ of infinite multiplicity. The point $z=1$ is an
eigenvalue of multiplicity 1. \rm

{\bf Proof.}
Since $\|\uex\|=1$ we have $\spec \subseteq \D(1)$. Let
$z\in D(1)$.  Consider the Koopman operator of the cusp
family $V_\epsilon:X\to X$
$$
V_\epsilon f(x)=f(F_\epsilon(x)).
$$
One can directly verify that the functions $\psi$,
\beq
\psi(x)=\sum_{k=0}^\infty z^kV_\epsilon^k h,
\label {uvuv}
\eeq
where $h(x)=g(x)(1+2\epsilon x)$ and $g(x)$ is an odd function,
are eigenfunctions of $\ue$:
$\ue\psi=z\psi$. As $g$ is an arbitrary odd function, this
proves that all points of $D(1)$ are eigenvalues of $\uex$
of infinite multiplicity. $\Box$

{\bf Remark 1.} Formula (\ref{uvuv}) provides all the eigenfunctions of 
$\uex$ with eigenvalue $z$.

{\bf Remark~2.} \ Proposition~1 and its proof remain valid for the
Frobenius--Perron operator $U$ of any continuous exact endomorphism 
(instead of $h$ one should take any element of ker$\,U$).

{\bf Proposition 2.} \sl Let $\epsilon\in[0,1/2]$,
$n=1,2,\dots$, $X$ be the Banach space $C^n[-1,1]$.
Then the spectrum $\spec$ contains the closed disk
$\D(0,(1/2+\epsilon)^{n+1})$, and any point of the open disk
$D(0,(1/2+\epsilon)^{n+1})$ is a $($non-normal$)$
eigenvalue of $\uex$ of infinite multiplicity. The set
$S=\spec\setminus\D(0,(1/2+\epsilon)^{n})$ is finite and
any $z\in S$ is a normal eigenvalue of $\uex$. \rm

{\bf Remark 3.} \ Under the conditions of
Proposition~2
for any $z\in S$ and any $f\in\E(z,\uex)$ the function $f$
admits the analytic continuation to the disk
$D(0,1/\epsilon-1)$ if $\epsilon\in(0,1/2)$ and to the whole complex 
plane if $\epsilon=0$. This can be proved by estimating the
growth of the sequence $\displaystyle s_n=\sup_{t\in[-1,1]}
|f^{(n)}(t)|$.

{\bf Corollary 1.} \sl Let $\epsilon\in[0,1/2)$,
$X=C^\infty[-1,1]$ with the natural topology {\rm \cite{rob}}.
Then the spectrum $\spec$ is either finite or countable,
$0\in\spec$ and any point $z\in\spec\setminus\{0\}$ is a
normal $($and therefore isolated$)$ eigenvalue of $\uex$. 
Moreover for any
$z\in\spec\setminus\{0\}$ and any $f\in\E(z,\uex)$ the
function $f$ admits the analytic continuation to the disk
$D(0,1/\epsilon-1)$ if $\epsilon\in(0,1/2)$ and to the
whole complex plane if $\epsilon=0$.  \rm

{\bf Corollary 2.} \sl Let $n=1,2,\dots,\infty$ and $X$ be
the space $C^n[-1,1]$. Then the spectrum
$\sigma(U_{1/2}^X)$ is the closed unit disk $\D(0,1)$, and
any point of the open unit disk $D(0,1)$ is an eigenvalue
of $U_{1/2}^X$ of infinite multiplicity. \rm

{\bf Proof of Proposition 2.}

Let us define the sequence $t_n$ by the formula
\beq
t_0=1, \quad t_{n+1}=-a_\epsilon(t_n), \ \ n=1,2,\dots
\eeq
It is easy to see that $t_1=0$, the sequence $t_n$ is strictly
decreasing and
\beq
\begin{array}{l}
t_n= -1+\frac4n+O\Bigl(\frac1{n^2}\Bigr)\hbox{\ \ for\ \ }
\epsilon=1/2,
\\
t_n= -1+c(\epsilon)\Bigl(\frac12+\epsilon\Bigr)^n+
O\biggl(\Bigl(\frac12+\epsilon\Bigr)^{2n}\biggr),\hbox{\ \
for\ \ } \epsilon\in[0,1/2).
\end{array}
\eeq

Let $z\in\C$.
Pick an arbitrary function $\phi:(0,1]\to\C$.
Define recurrently the function $f_\phi:(-1,1]\to\C$ as
follows
\beq
\begin{array}{l}
f_\phi(x)=\phi(x) \hbox{\ \ for\ \ } x\in(0,1]=(t_1,t_0],
\\
f_\phi(x)=\frac{2zf(a_\epsilon^{-1}(-x))}{1-2\epsilon x}
-\frac{1+2\epsilon x}{1-2\epsilon x}f(-x)\hbox{\ \ for\ \ }
x\in(t_{n+1},t_n], \ n=1,2,\dots \end{array}
\eeq
It is straightforward to see that $f_\phi$ is the unique
function $f:(-1,1]\to\C$ for which $f
\bigr|_{(0,1]}=\phi$
and $\uex f(x)=zf(x)$ for all $x\in(-1,1]$. Let now $\phi$
be an element of $C^{\infty}[0,1]$ such that the support
of $\phi$ (i.e. the closure in [0,1] of the set
$\{t:\phi(t)\neq0\}$) is contained in the interval
$(0,-t_2)$.  It is clear that
$f_\phi\in C^\infty(-1,1]$. Using formula (11) and the
asymptotics (10), for any $z\in\C$, $|z|<(1/2+\epsilon)^{n+1}$
one can verify that
\beq
\lim_{t\downarrow-1}f_\phi^{(j)}(t)=0, \ \
j=0,1,\dots,n.
\eeq
Therefore putting $f_\phi(-1)=0$, we
see that $f_\phi\in X=C^n[-1,1]$ and $\uex f_\phi=zf_\phi$.
Hence $\spec$ contains
$\D(n+1,\epsilon)$ and any point of $D(n+1,\epsilon)$ is
an eigenvalue of $\uex$ of infinite multiplicity.

The second part of Proposition~2 follows from Ruelle's
results on spectra of positive transfer operators, see
\cite {baladi}, Theorem~2.5 and Exercise~2.9.

\section {Spectral properties of the operator $\ue$ in
spaces of analytic functions.}

The spectral properties of the operator $\ue$ in
spaces of analytic functions differ considerably depending
on the choise of the space and on the values $\epsilon=1/2$ or
$\epsilon \neq 1/2$. Furthermore, not all of these
properties are known yet.

{\bf Proposition 3.} \sl Let $\epsilon\in(0,1/2)$,
$q\in(1,1/\epsilon-1)$ and $X$ be the
Hardy space $\H^2(0,q)$.  Then the operator $\uex$ is
nuclear.  Moreover,
\begin{itemize}
\item[{\rm (i)}]
the eigenvalues $z_n=z_n(\uex)$ and the
eigenspaces $\E(z_n(\uex),\uex)$ do not depend on $q$;
\item[{\rm (ii)}]
the eigenvalues $z_n$ satisfy the inequality
\end{itemize}
\beq
|z_n|\leq1.5 c^n, {\sl \ \ where\ \ }
c = c(\epsilon)=\sqrt{1/2+\sqrt{\epsilon(1-\epsilon)}} < 1.
\label{ubound}
\eeq
\rm

{\bf Proof.}

It is easy to show that for any $r>1$
$$
\alpha(r)=\sup_{|z|=r}|a_\epsilon(z)|=
\frac12(1-\epsilon+r+\epsilon r^2).
$$
The function $\alpha$ is continuous and strictly increasing on 
the interval $(1,1/\epsilon-1)$, and
$\alpha(r)<r$ for any $r\in(1,1/\epsilon-1)$.  Put
$q'=\alpha^{-1}(q)>q$. From the definition of the operator
$\ue$ (\ref{froben}), it follows that $\ue$ is a linear
continuous operator from $\H^2(0,q)$ to $\H^2(0,q')$ with norm
less than or equal to $1+\epsilon\alpha(q)$. Thus $\uex$ is the 
composition of this operator and of the operator defining the 
embedding of $\H^2(0,q')$
into $\H^2(0,q)$, which is nuclear with $s$-numbers
$(q/q')^n$. Therefore, the operator $\uex$ is nuclear with
$s$-numbers $s_n(\uex)\leq(1+\epsilon\alpha(q))(q/q')^n$.

From Remark~3 follows that the eigenvalues
$z_n=z_n(\uex)$ and the eigenspaces
$\E(z_n(\uex),\uex)$ do not depend on $q$ and
coincide with the eigenvalues and eigenspaces of
$\ue^{C^\infty[-1,1]}$.

From Weyl's inequality \cite {weyl} we have
$$
|z_n|^{n+1}\leq\prod_{k=0}^n|z_k|\leq \prod_{k=0}^n
s_k(\uex)\leq (1+\epsilon\alpha(q))^{n+1}(q/q')^{(n+1)n/2}
$$
and therefore
\beq
z_n\leq(1+\epsilon\alpha(q))(q/q')^{n/2}.
\label {www}
\eeq
The ratio $q/q'$ is minimal for $q'=\sqrt{1/\epsilon-1}$ and is
equal to $1/2+\sqrt{\epsilon(1-\epsilon)}$. For this value of
$q'$ we have $1+\epsilon\alpha(q)\leq
1+\sqrt{\epsilon-\epsilon^2}\leq1.5$. Therefore
inequality (\ref{www}) for $q'=\sqrt{1/\epsilon-1}$ implies
(\ref{ubound}).  $\Box$

The case $\epsilon=1/2$ is much more difficult and so far
there exist very few results on the spectral properties of the 
Frobenius--Perron operators of the maps with parabolic
neutral fixed points. We would like to point out the result of
H.~Rugh \cite{rugh1}, who considered the Frobenius--Perron
operators of piece-wise analytic maps, which are expanding
everywhere except one parabolic fixed point. Namely, he
constructed a specific map-dependent Banach space of
analytic functions, where the spectrum of the F.P.O
consists of the segment [0,1] and some isolated
normal eigenvalues. This space is in fact the image of
$L_1[0,+\infty)$ with respect to some map-dependent integral transformation
(similar to the Laplace transform). This idea
applied to the cusp map allows to verify that the F.P.O.
$U_{1/2}$ has similar spectral properties
in certain weighted Hardy spaces in disks $D(\alpha,1+\alpha)$,
$0<\alpha<1$.

The result of H.~Rugh is very interesting since it provides
the first example of a Banach space of smooth functions,
where the spectrum of the Frobenius--Perron operator of the
cusp map is non-trivial. Note that the functions of
Rugh's space are analytic in all points of the
segment except the parabolic fixed point ($-1$ in our
case). However we should notice that the spectrum of the
F.P.O. of a map $S$ in spaces of
analytic functions with singularity at a fixed point of $S$
may differ considerably from the spectrum in spaces of
everywhere analytic functions. We illustrate this statement
for the simplest expanding map $F_0$, which is the tent map.

{\bf Proposition 4.} \sl Let $p\in[1,+\infty]$,
$0<\alpha<1$ and $X=\H^p(\alpha,1+\alpha)$. Then
the spectrum $\sigma(U_{0}^X)$ depends on $p$. Namely,
$\sigma(U_{0}^X)$ is the union
of the disk $\D(0,2^{2/p-1})$ and some set of
$($isolated$)$ normal eigenvalues.  \rm

{\bf Proof.} Evidently, $U_0^X=A+B$, where $\displaystyle Af(x)=\frac12f
\left(\frac{1-x}2\right)$, $\displaystyle Bf(x)=\frac12f
\left(\frac{x-1}2\right)$. Since the  image of the operator
$B$ is contained in the space $H^q(\alpha,\beta)$, where
$\beta=\min\{1+5\alpha,3-\alpha\}>1+\alpha$, the operator
$B$ is nuclear and therefore compact.

Let us estimate now the norm of the operator $A$. Let $f\in X$. Then
$$
\begin{array}{l}
\displaystyle
\|Af\|^q=\int\limits_{D(\alpha,1+\alpha)}\left(\frac12\left|f\left(
\frac{x+iy-1}2\right)\right|\right)^qdxdy=\int\limits_{D(\alpha-1,
(1+\alpha)/2)}\frac4{2^q}|f(x+iy)|^q\,dxdy\leq
\\
\displaystyle
\frac1{2^{q-2}}\int\limits_{D(\alpha,1+\alpha)}|f|^q\,
dxdy=\frac1{2^{q-2}}\|f\|^q.
\end{array}
$$

Therefore $\|A\|\leq 2^{\frac2q-1}$. On the other hand one can verify that $Af_\lambda=2^{-1-\lambda}f_\lambda$, where $f_\lambda(x)=(x+1)^\lambda$ and $f_\lambda\in X$ if and only if
$$
{\rm Re}\,\lambda>-\frac2q\iff |2^{-1-\lambda}|\leq 2^{\frac2q-1}.
$$

Hence, the open disk $D(0,2^{\frac2q-1})$ is contained in
the spectrum of $A$. Since $\|A\|\leq 2^{\frac2q-1}$, we
find that $\sigma(A)=\D(0,2^{\frac2q-1})$.

Since the operator $B$ is compact and $U_0^X=A+B$, the
theorem on holomorphic operator-functions (\cite {g-k}, Chapter I)
implies that the spectrum of $U_0^X$ is the union of
$\D(0,2^{\frac2q-1})$ and some (isolated) normal
eigenvalues.

{\bf Proposition 5.} \sl Let $0<\nu<0.3$ and
$X$ be the space of the
functions $f:(-1,1]\to\C$ such that the function
$g_f(z)=f(-1+2^{-z})$, $g:[-1,+\infty)\to\C$ admits the
analytic continuation to some element of the conventional Hardy Hilbert
space $Y$ in the half-plane $A_\nu=\{{\rm Re}\,z>-1-\nu\}$ $($We
transfer the scalar product from this Hardy space to
$X$ by the bijective linear transform $f\mapsto g_f)$.  Then
$\sigma(U_{0}^X)=[0,1]\cup S$, where $S$ consists of
normal eigenvalues. \rm

{\bf Remark 4} The space $X$ of Proposition~5 is a Hilbert space
of functions analytic on the set
$D(-1,c)\setminus(-1-c,-1]$ for some $c=c(\nu)>2$.

{\bf Proof of Proposition 5.}

From the definition of the scalar product in
$X$, the operator $T:X\to Y$, $Tf(x)=f(-1+2^{-x})$
is a unitary transformation. Therefore the operator
$$
W=TU_0T^{-1}:Y\to Y
$$
and $U_0$ are unitarily equivalent. From the definitions of $T$ and 
$U_0$ it follows that $W=A+B$, where
$$
Af(x)=\frac12 f(x+1),\qquad Bf(x)=\frac12 f(-\log_2(2+2^{-y-1})).
$$

It is straightforward to verify that the closure of the set
$\{-\log_2(2+2^{-y-1}):y\in A_\nu\}$ is a compact subset
of $A_\nu$. Hence, the operator $B$ is nuclear and therefore compact.
On the other hand the conventional Laplace transform and a linear change 
of variables provide a unitary equivalence between the operator $A$ and 
the operator of multiplication with the function $e^{-t}$ acting on a 
certain weighted Sobolev space of functions on $[0,+\infty)$. Therefore 
the spectrum of $A$ is the segment $[0,1]$.

Since the operator $B$ is compact, the theorem on holomorphic 
operator-functions~\cite {g-k} implies that the spectrum of
$U_0^X$, which is identical with the spectrum of $W$ is
the union of the segment $[0,1]$ and some set of (isolated)
normal eigenvalues.  $\Box$

It is worth noticing that the space constructed in
Proposition~6 is obtained by the method similar to the
construction of Rugh \cite{rugh1}. Thus, it is not a priori
clear what is the origin of the ``continuous spectrum''
[0,1] obtained in \cite{rugh1}: the dynamical properties of
the map or the choice of the space.

We conjecture that in the space of real-analytic on
$[-1,1]$ functions, the point spectrum of the
Frobenius--Perron operator $U_{1/2}$ of the cusp map is
$\{0,1\}$, i.e., the eigenfunction equation $U_{1/2}f=zf$ has
non-zero analytic solutions only for $z=0$ and $z=1$. To support 
this conjecture we show that $\{0,1\}$ is
the point spectrum of $U_{1/2}$ in the space of entire
functions.

{\bf Proposition 6.} \sl Let $\epsilon=1/2$, $X$ be the
space of entire functions. Then the spectrum of $\uex$
is the whole complex plane $\bf C$ and the point spectrum
of $\uex$ is the two-point set $\{0,1\}$. The eigenvalue 0
has infinite multiplicity, and the eigenvalue 1 has
multiplicity 1. \rm

{\bf Proof.}  The ergodicity of the map
$F_{1/2}$ implies the multiplicity 1 for the
eigenvalue $z=1$. The null space of the operator $U_{1/2}$ is
$$
\{f\in X:f(x)=(1+x)g(x):g\hbox{\ \ is an odd function.}\}
$$
Therefore 0 is an eigenvalue of $U_{1/2}$ of infinite
multiplicity. Let now $z\in\C\setminus\{0,1\}$, $\psi \in X$
and $U_{1/2} \psi=z\psi$.  The eigenvalue equation for $x=1$
implies that $\psi(-1)=0$. Therefore
$\psi(x)=(1+x)g(x)$ for some $g\in X$. Let $\xi(x)=g(x)+g(-x)$.
The eigenvalue equation $U_{1/2}\psi=z\psi$ in terms of the
function $\xi$ can be rewritten as
\beq
\xi \biggl(\Bigl(\frac{x+1}2\Bigr)^2\biggr)
=\frac{32z \xi(x)}{x^3+5x^2+11x+15}+
\frac{x^3-5x^2+11x-15}{x^3+5x^2+11x+15}
\xi\biggl(\Bigl(\frac{x-1}2\Bigr)^2\biggr).
\label{M}
\eeq
Let $\displaystyle M(R)=\max_{|x|=R}|\xi(x)|$,
$c\in(0,\sqrt2)$. It is easy to see that if $x\in\C$,
Re$\,(x+1)^2\geq0$, Re$\,x\geq0$, and
$R\leq|(x+1)^2/4|\leq R+c\sqrt R$ then, for sufficiently
large $R>0$, $|x|\leq R$ and $|(x-1)^2/4|\leq R$. Since $\xi$
is even this fact together with formula (\ref{M}) imply that
\beq
M(R+\sqrt R+1/4)\leq M(R)(1+5/\sqrt R+O(1/R))\quad {\rm when}
\quad (R\to +\infty).
\label{N}
\eeq
Applying (\ref{N}) to $R_n=n^2/4$ and using the equality
$R_{n+1}=R_n+\sqrt{R_n}+1/4$, we obtain
$$
M(n^2/4)\leq c_1\prod_{k=1}^n(1+10/k+O(1/k^2))
$$
for some positive constant $c_1$. Therefore $M(n^2/4)=O(n^{10})$,
and $M(R)=O(R^5)$. This estimation implies (see~\cite{complex})
that $\xi$ is a polynomial of degree at most 5. On the other hand,
using induction with respect to the degree of polynomial, one can
show that there are no polynomial solutions of the equation (\ref{M}).
$\Box$

{\bf Remark 5.} Similar technique allows verifying that for
any $z\in\C$, $z\neq0$ the function $f(x)=x$ does not
belong to the space $U_{1/2}(X)$, where $X$ is the set of
all entire functions. Therefore, the spectrum of
$U_{1/2}^X$ is the whole complex plane.

\section {Numerical results for the spectra}

In the previous section we presented the general description of the
spectrum of the operator $\ue$. However, the eigenvalues and the
eigenfunctions of the cusp family are not known explicitly.
So we should compute them numerically.
In order to perform this calculation in the space of the
analytic functions, we use Taylor's expansion:
\beq
f(x) = \sum_{k=0}^N c_k x^k.
\label{taylor}
\eeq
The eigenvalue problem $\ue f(x) =  z  f(x)$ can be reformulated in
terms of the coefficients $c_k$:
\beq
\ue f(x) = \sum_{k=0}^N c_k \ue x^k = \sum_{k=0}^N c_k
\sum_{p=0} a_{pk} x^p =  z  \sum_{k=0}^N c_k x^k.
\label{matrix}
\eeq
As the operator $\ue$ is nuclear, we can project the last expression
onto the subspace $\{x_k\}_{k=0}^{N}$. Now the eigenvalue problem
can be written as $A\vec{c} =  z \vec{c}$, where $\{A\}_{kp}=a_{kp}$,
see (\ref{matrix}).

The coefficients $a_{pk}$ in~(\ref{matrix}) are equal to
\beq
a_{pk} = \left\{
\begin{array}{ll} (-1)^p
f(\epsilon,k,p), & k \quad  \mbox{is even,}  \\
(-1)^{p+1}2\epsilon f(\epsilon,k+1,p), & k \quad \mbox{is odd,}
\end{array} \right.
\label{acoef}
\eeq
where the function $f(\epsilon,k,p)$ is defined as
$$
f(\epsilon,k,p) = {1\over 2^k} \sum_{l=0}^p C_k^l C_k^{p-l}
\epsilon^l (1-\epsilon)^{k-l}.
$$
The most precise and convenient way for the calculation of the
coefficients $a_{pk}$ is the use of the recurrence relation:
$$
a_{p,k+2} = {1\over 4} \left \{ (1-\epsilon)^2 a_{p,k} +
(2\epsilon-2)a_{p-1,k} +(-2\epsilon^2+2\epsilon+1)a_{p-2,k}
-2\epsilon a_{p-3,k}+\epsilon^2 a_{p-4,k} \right \}.
$$
This representation is much more accurate than the numerical integration
used in~\cite {kaufmann} hence it permits using longer
expansion~(\ref{taylor}) without loss of accuracy.

It is worth noticing that the matrix $A$ in non-symmetric. Up to
$2 \times 10^3$ terms in the expansion~(\ref{taylor}) were used to
get converged results. In order to check convergence, we use the trace
formula for the operator $\ue$. Namely, 
as for $\epsilon \in [0,1/2)$ the operator $\ue$ is nuclear,
we can calculate its trace by using the Grothendieck-Fredholm formula
(see for example~\cite {baladi,cvitanovic}):
\beq
\hbox{tr}\, \ue = \sum_{n=0}^\infty z_n =
{1 \over 1/2-\epsilon } - {2 \over \sqrt{9-4\epsilon(1-\epsilon)}}
\label {trace}
\eeq
and compare this value with the numerical calculations.

In Fig.~1, ten maximal eigenvalues of the operator $\ue$ are presented.
Because of very good convergence of our numerical method for small
$\epsilon$, the asymptotics of the $z_n$ as $\epsilon\rightarrow 0$
can be numerically calculated:
\beq
{z_{n+1} \over z_n}=\frac{1}{4}+
\left(2n-\frac{1}{2}\right)\epsilon+O(\epsilon^2).
\label {ratio}
\eeq
Hence the cusp family has neither spectrum in the form $r^n$, where 
$n \in \N$, $r \in \R$, nor combination of few such spectra when 
$\epsilon\neq 0$.

Using relation~(\ref{ratio}), we can find a general formula for the
eigenvalues when $\epsilon$ is small:
\beq
z_{n+1} = \left(1\over 4 \right)^n (1+2n(2n+1)\epsilon + O(\epsilon^2)),
\quad n=0,1,2,\ldots .
\label{evasymp}
\eeq
This result gives for the asymptotics of the trace
\beq
tr \ue = {4\over 3} + {104\over 27}\epsilon +O(\epsilon^2)
\quad \mbox{when}\quad \epsilon\rightarrow 0.
\label{zeroas}
\eeq
Formula~(\ref{zeroas})
coincides with the asymptotics of Eq.~(\ref{trace}).
This coincidence supports strongly formulas~(\ref{ratio},\ref{evasymp})
which are obtained only numerically.

When $\epsilon\rightarrow 1/2$ and $n$ is fixed,
one can see that $z_n \rightarrow 1$. This result agrees with the
divergence of the trace. We have also checked that the eigenvalues
have the asymptotics
\beq
z_{n} = (1/2+\epsilon)^n \quad \mbox {when} \quad
\epsilon \to 1/2,
\eeq
that agrees with the asymptotics found in~\cite {kaufmann}.

Let us now discuss the eigenfunction behavior. In Figs.~2a, 2b we present
the second and fourth eigenfunctions, respectively, for few values of
$\epsilon$. One can easily see a concentration effect in a vicinity of
$-1$ as
$\epsilon\rightarrow 1/2$. The eigenfunctions tend to have the support only
at the point $x=-1$. This behavior is in a good agreement with the existence
of a ``formal eigenfunction" $\delta(x+1)$ for $\epsilon=0.5$. Such
behavior of eigenfunctions supports numerically the conjecture about the
non-existence of non-trivial (except of $\{0,1\}$) spectrum for the cusp
map in the space of the real analytic functions as the limit functions
have a singularity at the point $-1$.

\section {Conclusions}

The spectral properties of the cusp family~(\ref{familydefinition}) that
``interpolates'' between the tent map and the cusp map have
been investigated in different function spaces. This study has
permitted us to formulate a conjecture about the spectrum of the cusp map.
While some results about this spectrum can be proved, the general
description in different spaces of analytic functions is still unknown.

There are few questions which are particularly interesting in this context.
First, the question about the asymptotics of the autocorrelation
function for the cusp map. As the resonance eigenvalues tend to unity, one
can expect non-exponential decrease of the autocorrelation function.
The estimations in paper~\cite {correl} show that the autocorrelation
function $C(n)$ decreases as $1/n$ when $n\to\infty$. However, this
conjecture is not yet analytically proven. Another question addresses
the choice of the space of analytic functions where the spectrum of
the F.P.O. is naturally defined by the dynamics of the map.
Moreover, our calculations and calculations of~\cite {kaufmann} show
that the spectrum of the cusp family is real. While there are some
analytical results about a reality of the spectrum~\cite {rugh2}, they
are not applicable to the cusp family. Hence the question about the
reality of the spectrum also remains open.

\vskip 1cm

{\bf Acknowledgments.} We would like to thank Profs. Ilya Prigogine and
Victor Sadovnichy for helpful discussions. This work
enjoyed the financial support of the  European Commission,
project IST-2000-26016 IMCOMP, and the Belgian Government through the
Interuniversity Attraction Poles. S.~A.~Shkarin is
supported by the Alexander von Humboldt foundation.

\newpage
\vskip 1cm
\begin {center}{\bf Figure captions} \end {center}

Fig. 1. Ten maximal eigenvalues as the functions of
$\epsilon$.

Fig 2a, 2b. The second (2a) and fourth (2b) eigenfunctions
for $\epsilon=$ 0 (the solid line), 0.25 (the long-dashed
line), 0.4 (the dashed line), and 0.48 (the short-dashed
line).

\end {document}
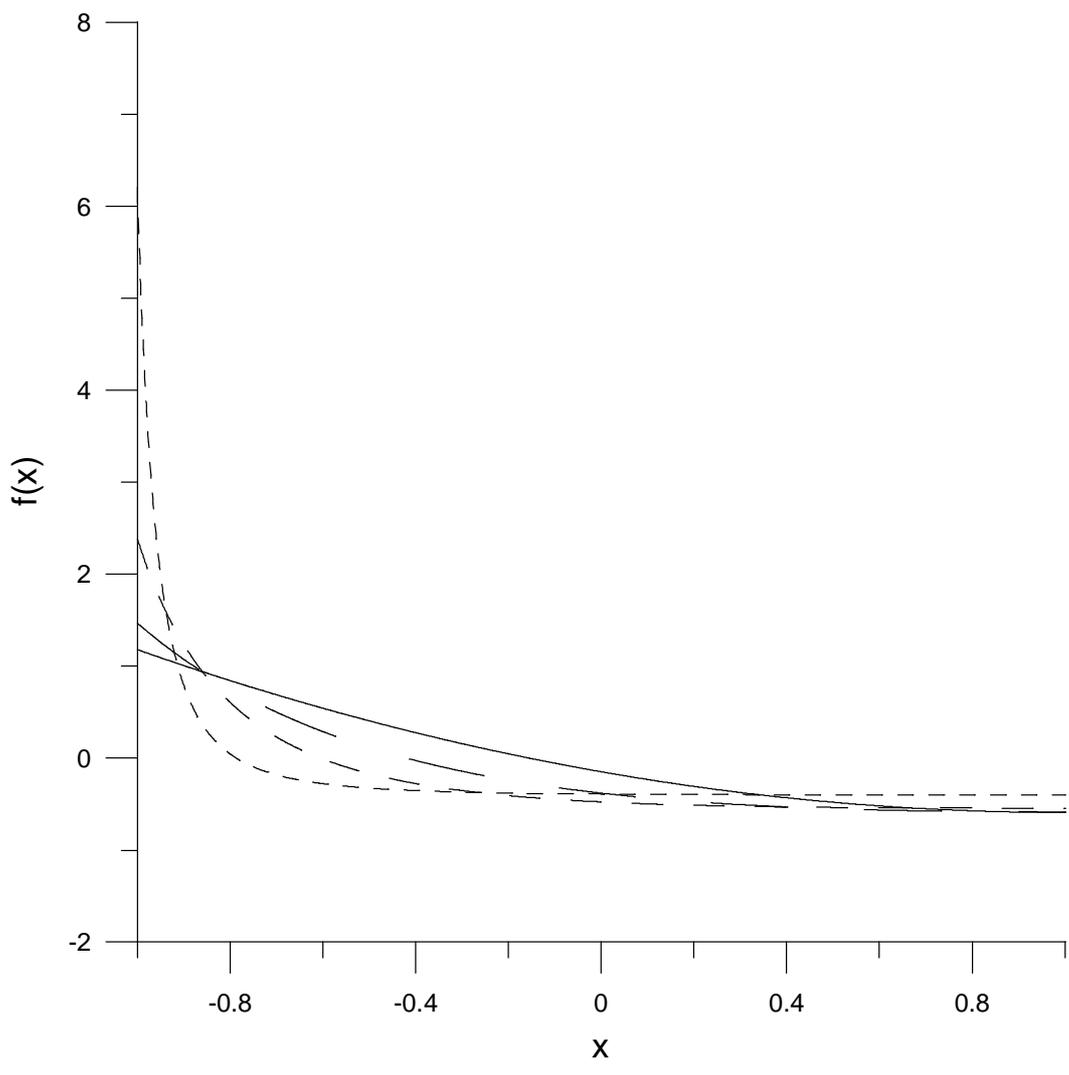
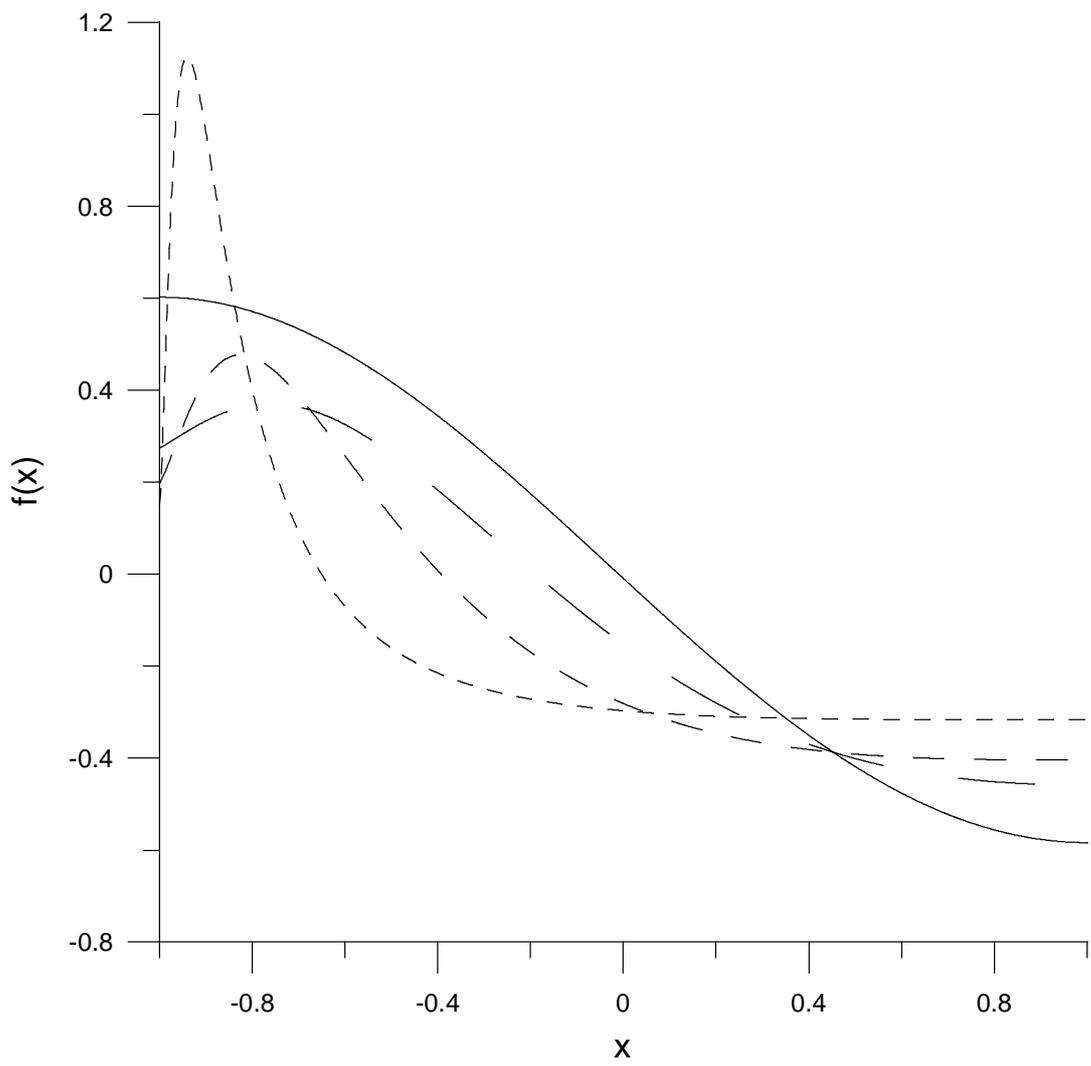
